\documentclass[reprint,amsmath,amssymb,noeprint,superscriptaddress,longbibliography]{revtex4-2}

\usepackage[english]{babel}
\usepackage[T1]{fontenc}
\usepackage[utf8]{inputenc}
\usepackage[colorinlistoftodos, color=green!40, prependcaption]{todonotes}
\usepackage{graphicx}
\usepackage{yhmath}
\usepackage{subfigure}
\usepackage{mathdots}
\usepackage{MnSymbol}
\usepackage{dsfont}
\usepackage{soul}
\usepackage[normalem]{ulem}
\usepackage{hyperref}
\usepackage{mathtools}
\usepackage{makecell}

\hypersetup{
    unicode=true, 
    plainpages=false,
    colorlinks=true,
    linkcolor=blue,
    citecolor=blue,
    filecolor=blue,
    urlcolor=blue
}

\usepackage{mathtools}

\DeclarePairedDelimiterX\braket[2]{\langle}{\rangle}{#1\,\delimsize\vert\,\mathopen{}#2}

\newcommand{\tphi}{\tilde{\phi}}
\newcommand{\tx}{\tilde{x}}
\newcommand{\tildet}{\tilde{t}}
\newcommand{\tE}{\tilde{E}}

\begin{document}
\title{Instantons and Rarefaction Pulses as  Pathways to Global Phase Coherence in Gain-Based Optical Networks}
\author{Richard Zhipeng Wang}
\author{Natalia~G.~Berloff}
\email[correspondence address: ]{N.G.Berloff@damtp.cam.ac.uk}
\affiliation{Department of Applied Mathematics and Theoretical Physics, University of Cambridge, Wilberforce Road, Cambridge CB3 0WA, United Kingdom\\}

\begin{abstract}
We investigate how to reliably remove unwanted global phase windings in gain-based optical oscillator networks, thereby ensuring convergence to the true synchronized configuration corresponding to the XY Hamiltonian's global minimum. Focusing on one-dimensional rings and two-dimensional toroidal lattices, we show that two key strategies greatly enhance the probability of reaching the defect-free state. First, operating at a low effective injection rate just above threshold, exploits the amplitude degree of freedom, allowing the system to form transient zero-amplitude holes: instantons in one dimension, or vortex-antivortex rarefaction pulses in two-dimensional space, that enable phase slips. Second, preparing the initial condition with amplitude or phase inhomogeneities can directly seed these amplitude collapses and prevent the system from getting trapped in higher-energy states with nonzero winding. Using the Stuart--Landau and Ginzburg--Landau equations as models for fast-reservoir lasers, we derive analytic and numerical evidence that even relatively minor amplitude dips can trigger global unwinding. We further demonstrate that the pump's slow annealing favours these amplitude-driven events, leading to improved success in finding the globally coherent ground state. These findings highlight the critical role of amplitude freedom in analogue solvers for XY optimization problems, showing how local amplitude suppression provides a direct route to ejecting topological defects.
\end{abstract}

\maketitle

\section{Introduction}
\label{sec:intro}

There is a growing interest in leveraging physics-inspired computing platforms based on networks of oscillators, including laser arrays, polariton condensates, and other nonlinear systems, to tackle complex optimization and dynamical problems \cite{pierangeli_large-scale_2019, pierangeli_adiabatic_2020, pierangeliScalableSpinGlassOptical2021a, kalininGlobalOptimizationSpin2018, kalininNetworksNonequilibriumCondensates2018, kalininPolaritonicNetworkParadigm2019, inagakiCoherentIsingMachine2016, yamamotoCoherentIsingMachines2017, honjo100000spinCoherentIsing2021, berloff2017realizing}. Instead of relying solely on conventional digital algorithms, these analogue computing approaches exploit the collective nonlinear dynamics of physical devices. Each oscillator in such networks can act as a “computational bit,” whose phase and amplitude evolve under mutual coupling in a manner that naturally explores the energy landscape of a target Hamiltonian. Systems ranging from degenerate optical parametric oscillators and polariton condensates to metallic nanolaser networks have been demonstrated to minimize effective spin Hamiltonians (e.g., Ising or XY), guiding the network into phase-locked steady states that correspond to low-energy solutions \cite{inagakiCoherentIsingMachine2016, kalininGlobalOptimizationSpin2018, partoRealizingSpinHamiltonians2020a}. Since many \(\mathrm{NP}\)-hard problems can be mapped onto XY- or Ising-type models with only polynomial overhead \cite{lucas_ising_2014}, these oscillator-based physical solvers offer a novel and promising route to exploring complex optimization tasks.

Beyond potential computing applications, coupled oscillator systems also serve as paradigmatic models of emergent collective behaviour in physics, biology, and engineering \cite{williams2013synchronization, denes2021synchronization}. Their universal features, such as phase locking, synchronization transitions, defect formation, and pattern selection, capture essential elements of coordination in real-world networks, from power grids to neuronal circuits. In particular, gain-dissipative oscillator arrays such as class-B lasers or exciton-polariton condensates mimic the rich phenomenology of nonlinear media and have been used to observe spontaneous synchronization, vortex formation, and other nonequilibrium phase transitions \cite{cataliottiJosephsonJunctionArrays2001, ankerNonlinearSelfTrappingMatter2005, cormanQuenchInducedSupercurrentsAnnular2014, esqueda2023route}. Their dynamical behaviours often go far beyond what simpler phase-only (Kuramoto-like) models can capture, underscoring the importance of retaining both amplitude and phase variables to understand phenomena relying on the amplitude-induced defect healing.

Among the many possible network geometries, the ring topology (periodic boundary conditions) provides a particularly clean and symmetric platform for studying synchronization and topological effects \cite{denes2021synchronization, esqueda2023route}. A ring of \(N\) oscillators eliminates boundary inhomogeneities and allows for uniform coupling around a closed loop. Despite its conceptual simplicity, such a ring can sustain a wealth of nontrivial phase patterns, ranging from the globally in-phase state to travelling waves and {\it splays (twisted) states} characterized by a nonzero winding number or topological charge. These winding states correspond to configurations in which the total phase around the loop increments by \(2\pi q\) (for integer \(q\)), thereby creating topological defects associated with that phase gradient. Such twisted states can be stable over a broad parameter range and may represent local minima in the system’s energy-like landscape, impeding convergence to the fully synchronized (lowest-energy) state.

A foundational theoretical tool for understanding ring oscillator synchronization is the Kuramoto phase model.
In the classic all-to-all version, each oscillator \(i\) is described by a single phase \(\theta_i(t)\) evolving according to
\[
\dot{\theta}_i = \omega_i + \frac{K}{N}\,\sum_{j=1}^N \sin\bigl(\theta_j - \theta_i\bigr),
\]
where \(\omega_i\) is the natural frequency of oscillator \(i\), \(K\) is the global coupling strength, and $N$ is the total number of oscillators in the network.
This model famously exhibits a transition from incoherence to synchronization when \(K\) surpasses a critical threshold.
On a ring with identical oscillators and only nearest-neighbour coupling, one modifies the coupling term to sum over the two neighbours of each oscillator; for instance,
\[
\dot{\theta}_i = \omega \;+\; \frac{K}{2}\,\Bigl[\sin\bigl(\theta_{i+1}-\theta_i\bigr) \;+\; \sin\bigl(\theta_{i-1}-\theta_i\bigr)\Bigr],
\]
where \(\theta_{i\pm1}\) refers to neighbors of \(\theta_i\) in a ring of size \(N\), and indices are taken modulo \(N\).
Despite this apparently simple one-dimensional topology, the ring Kuramoto system can sustain multiple phase-locked patterns such as the uniform in-phase solution, travelling waves, and splay states where the phase increments by a constant amount from one oscillator to the next.
Extensions of the basic model, such as adding a phase lag (Sakaguchi--Kuramoto), time delays, or frequency dispersion, further enrich the dynamical repertoire \cite{denes2021synchronization, aranson2002world}.
Nevertheless, since the Kuramoto model fixes the amplitude of each oscillator, it cannot capture mechanisms in which local oscillations vanish, enabling phase slips that unwind a topological defect.

Quantum many-body models provide another viewpoint.
The Bose--Hubbard model, for instance, describes bosons on a lattice with tunnelling and on-site interactions \cite{gupta2024bose}.
On a one-dimensional ring, it supports both a superfluid phase, in which each site has a well-defined coherent amplitude and phase, and a Mott insulator phase, in which phase coherence is lost.
At the mean-field level, this reduces to a discrete nonlinear Schr\"odinger equation that also exhibits ring-specific effects such as quantized winding in the condensate phase.
Josephson junction arrays and discrete atomic Bose-Einstein condensates in ring traps have experimentally realized these ideas, revealing how phase coherence and winding can arise or dissipate depending on tunnelling, interactions, and dissipation \cite{williams2013synchronization}.

More generally, amplitude-phase oscillator models such as the Stuart--Landau system and the complex Ginzburg--Landau equation go beyond the phase-only picture by allowing each oscillator's amplitude to vary \cite{aranson2002world}.
In rings of Stuart-Landau oscillators, splay states can undergo amplitude-driven instabilities, leading to phase slips or richer chaotic dynamics.
This amplitude degree of freedom is crucial for understanding how a topological defect can heal since it is often during an amplitude collapse that the phase becomes undefined and can jump by \(2\pi\).

\begin{figure}
    \centering
    \includegraphics[width=\columnwidth]{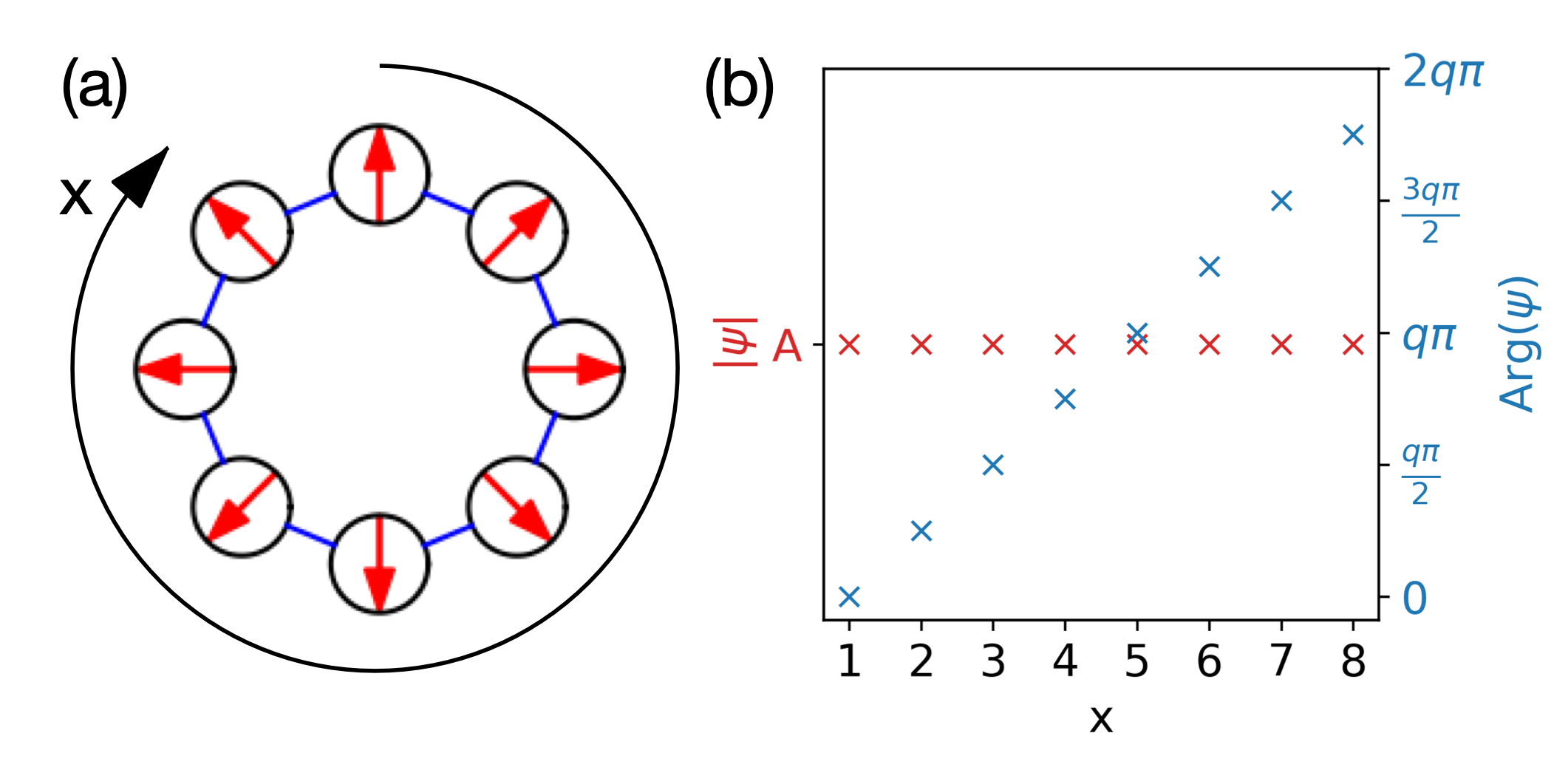}
    \caption{
    (a) A schematic diagram of a ring of $N$ identical lasers with uniform ferromagnetic nearest–neighbour coupling $\kappa>0$ (non‑zero only between adjacent sites); the coordinate $x$ indexes the lasers around the ring.
    (b) Steady state for $N=8$: amplitudes $|\psi|$ (\textcolor{red}{$\times$}) and phases $\arg\psi$ (\textcolor{blue}{+}) trace a uniform $2\pi q$ twist (shown for $q=1$).
    }
    \label{fig:ring_geometry}
\end{figure}

Returning specifically to lasers, a salient example is the class-B laser network, where each oscillator is governed by a complex field \(\psi_m\) and a real-valued gain \(G_m\).
The equations are
\begin{equation}
\frac{d\psi_m}{dt} \;=\; \frac{G_m - \alpha_m}{\tau_p}\,\psi_m \;+\; i\,\Omega_m\,\psi_m \;+\; \sum_{n}\,\frac{\kappa_{mn}}{\tau_p}\,\psi_n,
\label{eqn:laser_eqn_cmp}
\end{equation}
\begin{equation}
\frac{dG_m}{dt} \;=\; \frac{1}{\tau_c}\,\Bigl(P_m - G_m\,\bigl(|\psi_m|^2 + 1\bigr)\Bigr),
\label{eqn:reservoir_dynamics}
\end{equation}
where \(\alpha_m\), \(\Omega_m\), and \(\kappa_{mn}\) respectively denote the linear loss, detuning, and coupling strength, while \(\tau_p\) and \(\tau_c\) are characteristic timescales for photon and population dynamics, and \(P_m\) is the pumping rate.
In a ring geometry with uniform nearest-neighbour coupling (\(\kappa_{mn} = \kappa > 0\) for neighbours), these equations can admit steady-state solutions with nonzero winding number \(q\), referred to as phase-winding or twisted states as shown in Fig.~\ref{fig:ring_geometry}.
Although such configurations lie above the global minimum in energy, experiments and simulations in coupled laser arrays and polariton rings confirm that they can form spontaneously and persist  \cite{palObservingDissipativeTopological2017a, mahlerDynamicsDissipativeTopological2019, cooksonGeometricFrustrationPolygons2021, piccardoVortexLaserArrays2022}.

Experiments and simulations have shown that these phase-winding solutions do indeed occur in the context of laser and polariton networks. For instance, the investigation of the formation of topological defects in a one-dimensional ring of class-B lasers was performed in \cite{palObservingDissipativeTopological2017a, mahlerDynamicsDissipativeTopological2019}, while higher-charge phase windings in polariton condensates arranged in a ring were demonstrated in \cite{cooksonGeometricFrustrationPolygons2021}. In such gain-dissipative systems, any nonzero-winding steady state constitutes an excited state that lies above the global minimum in energy. However, the mechanism by which an initially prepared twisted (splay) configuration unwinds, reducing its winding number to zero and thereby reaching the fully synchronized ground state, depends crucially on the amplitude dynamics. In contrast, a purely phase-based Kuramoto model is often unable to heal topological defects, leaving a nonzero-winding state “stuck” if the phase slip events required for unwinding cannot occur \cite{palObservingDissipativeTopological2017a}.

Importantly, recent experimental works highlight that amplitude degrees of freedom can facilitate the decay of topological defects \cite{palObservingDissipativeTopological2017a, piccardoVortexLaserArrays2022}, but the exact reasons were not identified. We elucidate the process of unwinding and show that by allowing local oscillators to drop their amplitude temporarily, the system can create a phase slip event that changes the winding number---an effect absent in simpler constant amplitude or Kuramoto-like models. 

Connections to the Kibble--Zurek mechanism have also been proposed, identifying competing timescales for amplitude equilibration and phase ordering; if amplitude fluctuations remain “active” (like an effective thermal bath) longer than the time needed for phase information to spread, defects can be “pushed out” of the system \cite{palObservingDissipativeTopological2017a, cormanQuenchInducedSupercurrentsAnnular2014}.

Motivated by both fundamental and practical concerns, we, therefore, seek a clearer picture of how an initial splay state (i.e., a nonzero-winding configuration) can unwind to the globally synchronized state in ring oscillator networks. This question resonates across applications: in oscillator-based computing, one requires that any unwanted phase twist not trap the system in a local minimum, preventing the identification of a true ground-state solution \cite{stroev2023analog, cumminsIsingHamiltonian2025}. Similarly, controlling and eliminating phase defects is often essential for stable operation in coherent light generation or advanced photonic devices \cite{butow2024generating}.

\begin{figure}
    \centering
    \includegraphics[width=0.7\columnwidth]{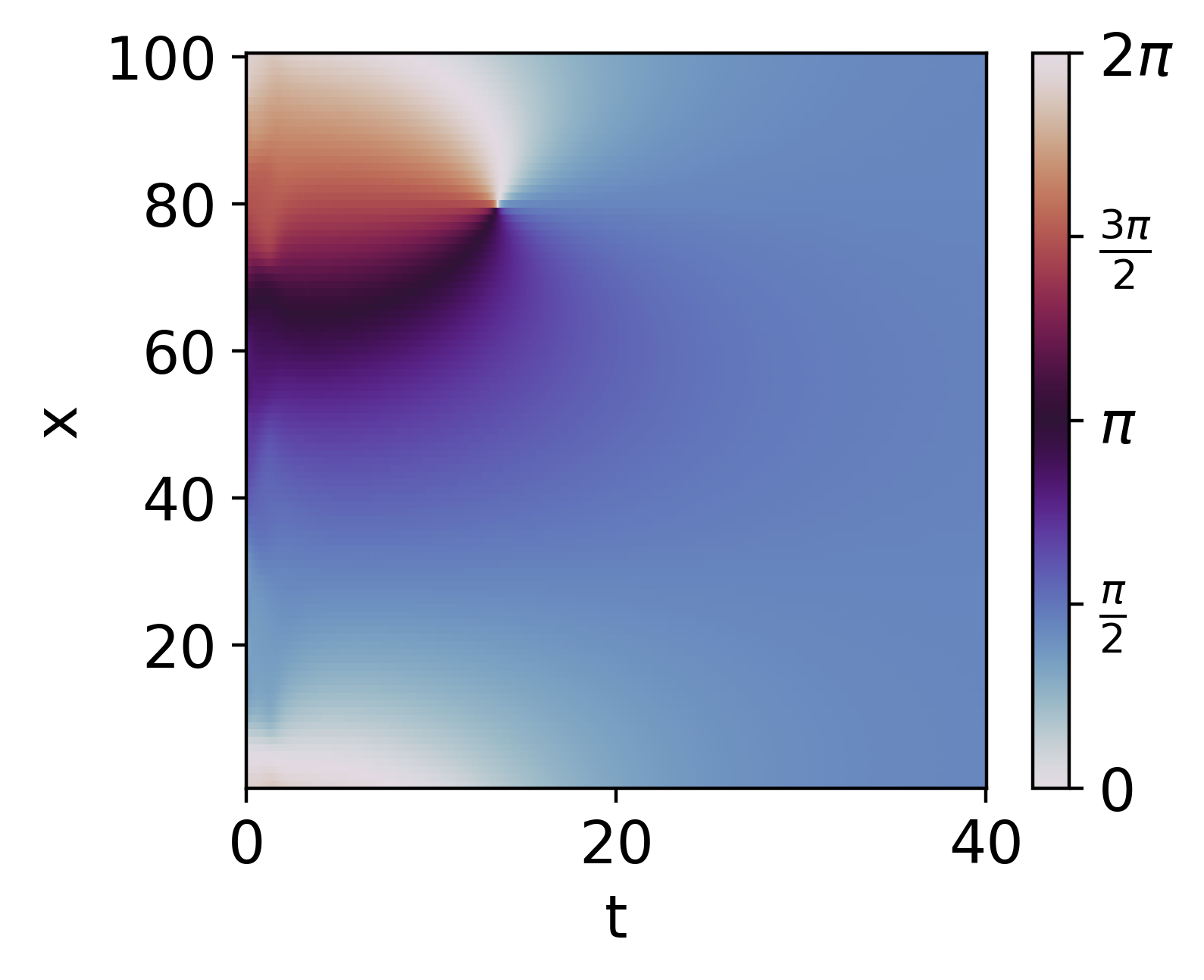}
    \caption{Space--time plot of the phases of a 100‑laser ring evolved with Eqs.~(\ref{eqn:laser_eqn_cmp})--(\ref{eqn:reservoir_dynamics}) using an adaptive \textsc{RK}4 scheme.  Parameters: $\tau_p = 10^{-3}$, $\tau_c = 50$, $\Omega_m = 0$, nearest‑neighbour $\kappa = 10^{-3}$, and $\alpha_m = 2.01\times 10^{-3}$.  The vertical axis is the site index, the horizontal axis is time, and hue encodes the phase modulo $2\pi$.  An instanton (vortex) nucleates at $t\!\approx\!17$, erasing the initial $2\pi$ winding and letting the system relax to the uniform ground state.  Initial amplitudes were $\approx 0$ with a $2\pi$ phase twist.%
}
    \label{fig:instanton_example}
\end{figure}

In this paper, we elucidate a mechanism by which topological defects dissipate in an optical ring network modelled by Stuart--Landau equations and laser rate equations. Our analysis bridges nonlinear dynamics and laser physics, clarifying how amplitude instabilities in the ring geometry can trigger phase slips (often referred to as instantons in the space-time diagram \cite{peninWhatBecomesVortices1996, nittaIncarnationsInstantons2014, nittaDecomposingInstantonsTwo2012, hananyVorticesInstantonsBranes2003}). 
As visualized in Fig.~\ref{fig:instanton_example}, these defects can nucleate in the amplitude degree of freedom, induce a discrete jump in winding number, and thereby allow the system to evolve into the zero-winding, fully synchronized ground state. Our focus is on systematically revealing the dynamical pathways and parameter regimes that enable such unwinding. By clarifying how phase gradients can be neutralized in a realistic model that retains amplitude dynamics, we provide insights relevant to the design of gain-based oscillators capable of escaping local minima. Ultimately, understanding this defect-healing process helps understand synchronization and the role of topology in complex oscillator networks, with implications ranging from fundamental studies of pattern formation in nonlinear media to practical implementations of oscillator-based computing.

Moreover, while much of our analysis focuses on a one-dimensional ring, in Section~\ref{sec:twod}, we show how these ideas naturally extend to two-dimensional toroidal geometries. In that setting, collision events involving vortex--antivortex pairs produce rarefaction pulses that are low-density fronts analogous to the 1D dark soliton \cite{jones1982motions}, that can span the entire domain and enable global $2\pi$ phase slips. This highlights a unifying role of amplitude freedom across different topologies and dimensions, reinforcing the conclusion that allowing local amplitude collapses is key to dissolving topological constraints in gain-based networks.

Our paper is arranged as follows. In Sec.~\ref{sec:analytics}, we present the mathematical model that features the formation of the instantons.
In Sec.~\ref{sec:numerical_results}, we provide numerical results demonstrating the process of instanton formation and the unwinding of topological defects.
Section ~\ref{sec:twod} addresses the 2D toroidal extension and explains how vortex--antivortex collisions enable analogous global phase slips in higher dimensions.
Finally, in Sec .~\ref {sec:discussion_conclusion}, we summarise our findings and discuss the implications of our results for optical hardware-based optimisation.

\section{Mathematical Models}
\label{sec:analytics}

\subsection{From Stuart--Landau Lattices to Continuum Ginzburg--Landau Equations}
\label{subsec:continuous_model}

In this subsection, we demonstrate how a discrete laser network in one or two dimensions reduces to a \emph{continuum} Ginzburg--Landau equation under smoothness assumptions.

We first consider an optical system with a \emph{fast} relaxing reservoir, such as class-A or class-C lasers, so that the reservoir equation~\eqref{eqn:reservoir_dynamics} is effectively in a steady-state regime.  Under this assumption, substituting the reservoir’s algebraic solution for $G_m$ back into the laser equations simplifies them to
\begin{equation}
    \frac{d\psi_m}{dt} 
    \;=\;
    \frac{1}{\tau_p}\,\Biggl(\frac{P_m}{|\psi_m|^2 + 1} - \alpha_m\Biggr)\,\psi_m
    \;+\;
    \sum_{n}\,\frac{\kappa_{mn}}{\tau_p}\,\psi_n,
\label{eqn:laser_eqn_fast_reservoir}
\end{equation}
where $i\,\Omega_m\,\psi_m$ in \eqref{eqn:laser_eqn_cmp} was absorbed into a rotating-frame phase for $\phi_m$, assuming uniform or small detuning.  We focus on a regime in which the pumping term
can be Taylor expanded to the lowest order in $|\psi_m|^2$, yielding the Stuart-Landau equation
\begin{equation}
    \frac{d\psi_m}{dt}
    \;=\;
    \biggl(\frac{P_m - \alpha_m}{\tau_p} \;-\; \frac{P_m}{\tau_p}\,|\psi_m|^2\biggr)\,\psi_m
    \;+\;
    \sum_{n}\,\frac{\kappa_{mn}}{\tau_p}\,\psi_n.
    \label{eqn:model_3}
\end{equation}
The saturable nonlinearity term $(P_m/\tau_p)\,|\psi_m|^2$ restricts unlimited amplitude growth even for large pumping $P_m$.
Defining
\begin{equation}
    \phi_m 
    \;=\;
    \sqrt{\frac{P_m}{\tau_p}}\;\psi_m,
    \quad
    \gamma 
    \;=\; 
    \frac{P_m - \alpha_m}{\tau_p},
    \quad
    J_{mn}
    \;=\;
    \frac{\kappa_{mn}}{\tau_p},
    \label{eqn:sl_conversion}
\end{equation}
we reduce the system to 
\begin{equation}
    \frac{d\phi_m}{dt} 
    \;=\;
    \bigl(\gamma - |\phi_m|^2\bigr)\,\phi_m
    \;+\;
    \sum_{n} J_{mn}\,\phi_n.
    \label{eqn:sl_complex}
\end{equation}
Here, $\gamma$ is the effective linear gain/loss parameter, and $J_{mn}$ encodes the oscillator coupling.    
Hence Eq.~(\ref{eqn:sl_complex}) captures the essential gain-saturation and coupling dynamics.

Because each oscillator’s phase $\theta_m$ in Eq.~\eqref{eqn:sl_complex} can be mapped onto an XY spin, this Stuart-Landau system can effectively minimise an XY-like Hamiltonian.  In particular, the ground state of the network with the uniform amplitude corresponds to the global minimum of an XY spin energy.  More concretely, interpreting $\phi_m = |\phi_m|\,e^{i\theta_m}$, we see that mutual coupling attempts to align phases $\{\theta_m\}$ across neighbouring sites as to minimize $H_{xy} = -\frac{1}{2}\sum_{n,m}^N cos(\theta_n-\theta_m)$, while the amplitude term $(\gamma - |\phi_m|^2)\phi_m$ saturates each oscillator at a near-constant magnitude.  Thus, when the system converges to its steady state under the gain-based dynamics, it is effectively “solving” the problem of synchronizing phases in an XY spin network, which is a framework that underlies several optical, laser and polaritonic Ising/XY machines proposed for combinatorial optimization \cite{inagakiCoherentIsingMachine2016,nixon2013observing,kalininGlobalOptimizationSpin2018, berloff2017realizing}.  In this sense, preventing (or annihilating) undesired phase windings becomes essential since any nontrivial winding is a higher-energy state of the XY Hamiltonian.  The following sections demonstrate precisely how amplitude freedom mediates the removal of such topological defects, thereby ensuring that the network can locate the correct (fully coherent) ground state.

{\it Nearest-neighbour couplings.}
In the subsequent analysis, we consider ferromagnetic nearest-neighbour couplings that approximate a discrete Laplacian on 1d or 2D lattices.  For a 1D ring of $N$ oscillators so that each oscillator is coupled to the nearest neighbours with $J_{nm}=l^{-2}>0,$ for a small real parameter $l>0$
with periodic boundary ($\phi_{N+1} \equiv \phi_1$).  In 2D, we consider an $N_x\times N_y$ square lattice with spacing $l$, where each site $(i,j)$ couples to its four neighbors $(i\pm1,j)$ and $(i,j\pm1)$ by $1/l^2$.  
We define a shifted gain parameter $\mu$ by:
\[
\mu 
\;=\;
\gamma 
\;+\;
\begin{cases}
  \tfrac{2}{l^2}, &\text{(1D)},\\
  \tfrac{4}{l^2}, &\text{(2D)}.
\end{cases}
\]
These definitions conveniently rewrite the coupling terms in a form reminiscent of $(\mu - |\phi|^2)\phi$ plus discrete second differences. To see this, regard $\phi_m(t)$ as $\phi(x_m,t)$ at $x_m = m\,l$.  Under smoothness assumptions,
\[
\phi_{m\pm1}(t)
\;\approx\;
\phi(x \pm l,t)
\;=\;
\phi(x,t)\,\pm\,l\,\partial_x\phi + \tfrac{l^2}{2}\,\partial^2_{xx}\phi + \cdots
\]
so that 
$
\phi_{m+1} - 2\phi_m + \phi_{m-1} 
\;\approx\; 
l^2\,\partial^2_{xx}\phi, 
$
 Eq.~\eqref{eqn:sl_complex} yields the 1D complex Ginzburg--Landau equation
\begin{equation}
    \frac{\partial \phi}{\partial t}
    \;=\;
    \bigl(\mu - |\phi|^2\bigr)\,\phi
    \;+\;
    \partial^2_{xx}\phi,
    \qquad
    x\in [0,L],
    \label{eqn:GLE-1d}
\end{equation}
where $L = N\,l$ is the total ring circumference.

Next, for a 2D square lattice of size $N_x\times N_y$, each site $(i,j)$ has four neighbors $(i\pm1,j)$ and $(i,j\pm1)$ with coupling $1/l^2$.  The discrete Stuart--Landau equation Eq.~\eqref{eqn:sl_complex} in continuous limit  then becomes
 the 2D complex Ginzburg--Landau equation
\begin{equation}
\partial_t \phi(x,y,t)
\;=\;
\bigl(\mu - |\phi|^2\bigr)\,\phi
\;+\;
\Delta \phi,
\label{eqn:continuous_pde}
\end{equation}
with $(x,y)\in [0,L_x]\times[0,L_y]$.
Here $L_x=N_x\,l$ and $L_y=N_y\,l$ with periodic boundary conditions to model a torus.

So in both one and two dimensions, nearest-neighbour Stuart--Landau lattices reduce in the continuum limit to Ginzburg--Landau equations Eq.~\eqref{eqn:continuous_pde}
with $\mu = \gamma + 2/l^2$ (1D) or $\mu = \gamma + 4/l^2$ (2D).   The continuum Ginzburg-Landau equations thus offer a tool for studying defect formation, annihilation, and the conditions under which the system converges to a globally synchronised or vortex-free state.

\subsection{Stationary Solutions of the Continuous Model in 1D}
\label{subsec:analytical_ansatz}
The first stationary solution that satisfies the periodic boundary conditions  is the constant amplitude solution of the form:
\begin{equation}
    \phi(x) = A e^{2\pi q i \frac{x}{L}}~,
    \label{eqn:const_amp_sol}
\end{equation}
where $q$ is an integer (topological charge).

\begin{figure}
    \centering
    \includegraphics[width=\columnwidth]{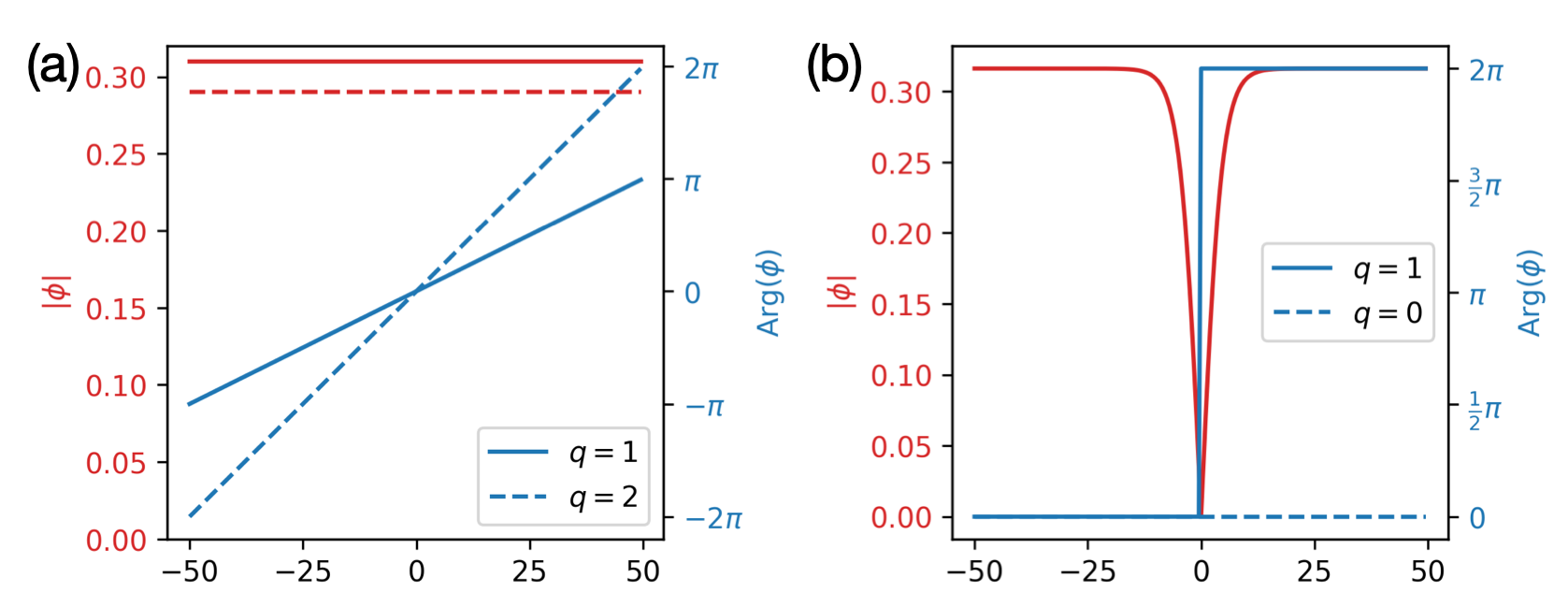}
    \caption{(a)~Constant‑amplitude steady states: amplitude $|\phi|$ and phase $\arg\phi$ for topological charges $q=1$ (solid) and $q=2$ (dashed). %
(b)~Instanton state: an amplitude dip to zero accompanied by two equivalent phase renderings—solid curve showing a $2\pi$ twist, dashed curve showing no net twist—demonstrating a continuous, zero‑energy path that links distinct winding sectors.%
}
    \label{fig:analytical_solutions}
\end{figure}

Substituting Eq.~\eqref{eqn:const_amp_sol} into Eq.~\eqref{eqn:continuous_pde}, we obtain a condition on $\mu$:
\begin{equation}
    \mu = A^2 + \left( \frac{2q\pi}{L} \right)^2 ~.
    \label{eqn:const_amp_mu}
\end{equation}
This implies that any $\mu < \left( \frac{2q\pi}{L} \right)^2$ cannot support a constant amplitude solution with topological charge $q$.

\begin{figure}
    \centering
    \includegraphics[width=0.7\columnwidth]{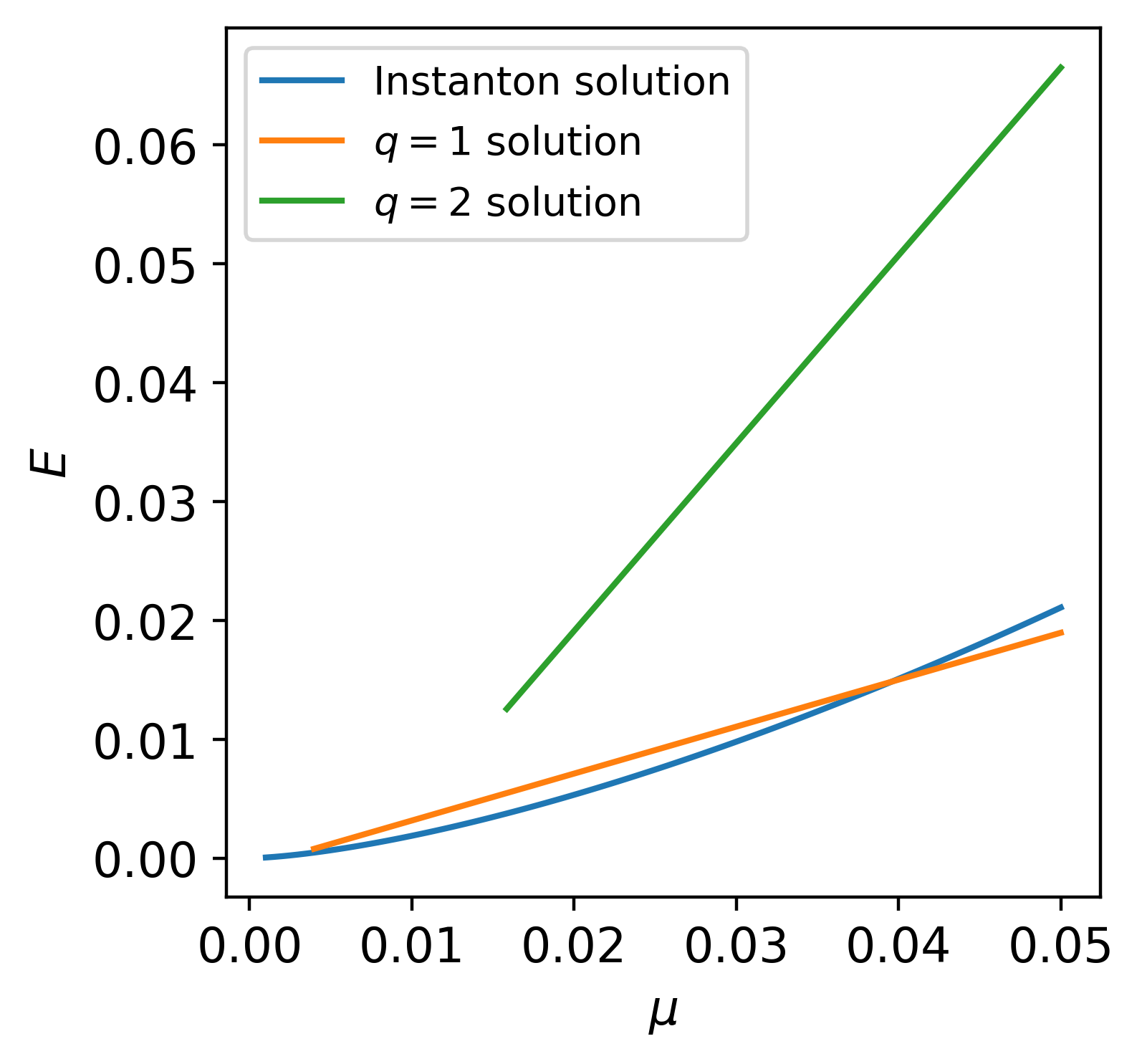}
    \caption{
    Analytic energy curves $E(\mu)$ for a constant -amplitude solutions for the ring of circumference $L=100$.  
Branches with fixed topological charge ($q=1,2$ shown) originate at their respective thresholds $\mu=(2\pi q/L)^2$, while the instanton branch exists for all $\mu>0$.%
}

    \label{fig:analytic_energy_vs_mu}
\end{figure}

If we consider Eq.~\eqref{eqn:continuous_pde} to be a gradient descent dynamics of the form
\begin{equation}
    \frac{\partial \phi(x,t)}{\partial t} = -\frac{\partial E}{\partial \phi^*}~,
    \label{eqn:gradient_descent}
\end{equation}
then the energy functional $E$ of the system is defined as
\begin{equation}
    E[\phi] = \int{\left| \frac{\partial \phi}{\partial x} \right|^2 + \frac{1}{2} \left( \mu - \left| \phi \right|^2 \right)^2}dx ~.
    \label{eqn:energy_functional}
\end{equation}
By substituting Eq.~\eqref{eqn:const_amp_sol} and Eq.~\eqref{eqn:const_amp_mu} into Eq.~\eqref{eqn:energy_functional}, the energy of the constant-amplitude stationary solution can be calculated analytically as:
\begin{equation}
    E = \left( \frac{2q\pi}{L} \right)^2 \left(\mu - \frac{1}{2}\left(\frac{2q\pi}{L}\right)^2 \right)L ~.
    \label{eqn:const_amp_energy}
\end{equation}
The energy $E$ is a linear function of $\mu$, and greater topological charge leads to solutions with higher energy at any given $\mu$, as shown in Fig.~\ref{fig:analytic_energy_vs_mu}.

\begin{figure*}[t]
    \centering
    \includegraphics[width=\linewidth]{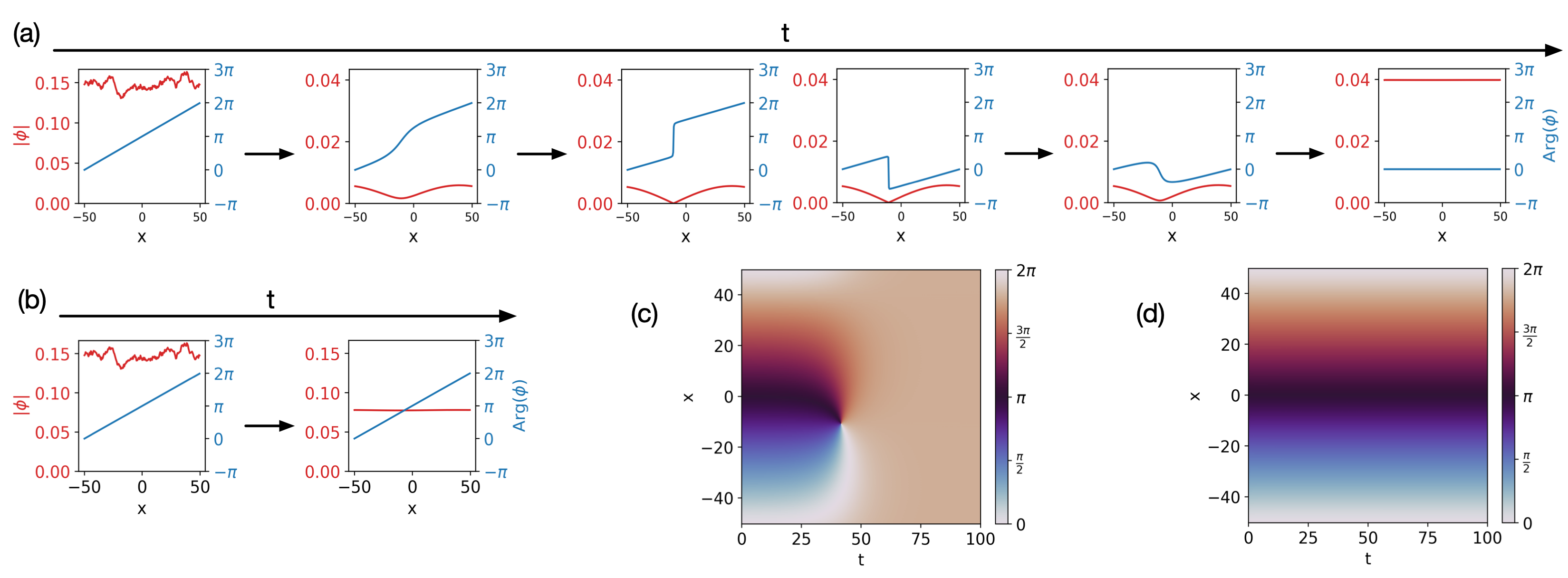}
    \caption{
    Instanton‐mediated unwinding vs.\ persistent winding in a 1‑D Ginzburg–Landau ring.  
(a)~Evolution of $|\phi|$ (top) and $\arg\phi$ (bottom) for $\mu = 0.002 < \mu^{\min}_{q=1}$: an instanton nucleates (local $|\phi|=0$ and $2\pi$ phase jump), erasing the initial twist and relaxing to the uniform ground state.  
(b)~Same initial state but $\mu = 0.01$; the $q=1$ branch is allowed, no instanton forms, and the system converges to the constant‑amplitude twisted solution.  
(c,d)~Space–time phase maps for cases (a,b); the instanton in (c) appears at $t \approx 40$, whereas the $2\pi$ winding persists in (d).  
Simulations integrate Eq.~\eqref{eqn:continuous_pde} on a ring of length $L=100$ (200 points) with adaptive 4\textsuperscript{th}‑order Runge–Kutta; initial amplitudes $|\phi|\in[0.1,0.2]$ (smoothed) and linear phase $\theta(x,0)=2\pi x/L$.%
}
    \label{fig:low_mu_evolution}
\end{figure*}

To construct a second stationary solution with $2 \pi$ winding in the periodic domain, we consider $\phi(x) = a(x) e^{i b(x)}$ with $b(x - L/2) = b(x+L/2) + 2 \pi$,  $a(0)=0$, and $a(x)$ approaches the constant as $x\rightarrow \pm L/2$.
Substituting into Eq.~(\ref{eqn:continuous_pde}) and separating real and imaginary parts gives
\begin{eqnarray}
  0&=& (\mu - a^2) a + a'' - a (b')^2, \nonumber\\
  0&=& 2 a' b' + b''. \label{2eq}
\end{eqnarray}
Since  $a(x)$ approaches a constant away from its depletion, $b(x)$ is a constant or a linear function there as follows from the second equation of Eqs.~(\ref{2eq}). If $b(x)$ is a constant function in these two limits, then to satisfy the boundary conditions on the winding, we get $b(x)=2\pi H(x),$ where $H(x)$ is the Heaviside step function. Combining this with the solution on $a(x)$ gives 
\begin{equation}
    \phi(x) = -\sqrt{\mu} \tanh{\left( \sqrt{\frac{\mu}{2}} x \right)} e^{i\pi H(x)}~. 
    \label{eqn:tanh_sol}
\end{equation}

Assuming $L$ is sufficiently large that $\phi(x)$ approaches a constant value everywhere far from $x=0$, the energy of this solution (Eq.~(\ref{eqn:energy_functional})) is given by:
\begin{equation}
    E = \frac{4\sqrt{2}}{3} \mu^{\frac{3}{2}} ~.
\end{equation}
Function $\phi(x)$ has constant phase everywhere except $x=0$ where it experiences a phase jump of $2\pi$, as shown in Fig.~\ref{fig:analytical_solutions}(b).
Hence, this solution can be considered as having a phase winding of $2\pi$.
However, it can also be equivalently considered as having no phase change at $x=0$ at all, and then it can be regarded as having no phase winding.
This means that this phase jump by $2\pi$ provides a pathway for a discrete topological charge to change from 1 to 0 without any abrupt jump in energy.
In other words, the topological defect with charge 1 degenerates with the globally coherent state with charge 0 at this point.
Hence, if the dynamics of the system passes through this point, then the system can unwind the initial phase winding by $2\pi$.
In order for phase of $\phi(x,t)$ at a point to jump abruptly(by $2\pi$ in this case) without going through any energy barrier, the amplitude of $\phi(x,t)$ at this point must be $0$ so that all phases are degenerate.
Indeed, this is the most important feature of this stationary solution - the occurrence of zero amplitude and $2\pi$ phase shift at the same point - a formation of an instanton.

In Sec.~\ref{subsec:examples_of_instanton_formation}, we present the numerical evidence of such an unwinding mechanism to decrease the non-zero topological charge of a random initial condition.

\subsection{Parametrised Initial Conditions}
To understand the basins of attraction of the above two stationary solutions, one can remove the parameter $\mu$ from the dynamical equation for the $\phi$ field (Eq.~\eqref{eqn:continuous_pde}) by defining:
\begin{equation}
    \begin{split}
        \tildet =& \mu t ~, \\
        \tx =& \sqrt{\mu} x ~, \\
        \tphi =& \phi / \sqrt{\mu} ~,
    \end{split}
\end{equation}
which reduces Eq.~\eqref{eqn:continuous_pde} to:
\begin{equation}
    \frac{\partial \tphi(\tx,\tildet)}{\partial \tildet } = (1 - |\tphi|^2)\tphi + \frac{\partial^2 \tphi}{\partial \tx^2} ~,
    \label{eqn:continuous_pde_scaled}
\end{equation}
and the equivalent energy functional is given by:
\begin{equation}
    \tE[\tphi] = E[\phi] / \mu^{\frac{3}{2}} ~.
\end{equation}

The advantage of considering Eq.~\eqref{eqn:continuous_pde_scaled} is that the result of the dynamical evolution now depends entirely on the initial conditions, namely the amplitude and phase profiles of the initial state, and does not depend on some arbitrary externally controlled parameter $\mu$ any more.

In numerical tests, we can now parametrise the initial condition $\tphi(\tx,\tildet=0) = A(\tx)e^{i\theta(\tx)}$:
\begin{equation}
\label{eqn:parameterised_amplitude}
    A(\tx) = \frac{e^{|\tx|} + d}{e^{|\tx|}+1} A_0 ~,
\end{equation}
where $-1 \le d \le 1$ is a parameter that determines how deep the drop of the amplitude at $\tx=0$ is and $A_0$ is a constant amplitude factor that will influence the energy $\tE[\tphi]$ of the initial state.
When $d=1$, the amplitude is constant for all $\tx$, and when $d=-1$, the amplitude drops to $0$ at $\tx=0$.
We also introduce a finite sized phase jump $\Delta \theta $ at $\tx=0$ parametrised by $\beta$
\begin{equation}
\label{eqn:parametrised_phase}
    \Delta \theta = \beta \pi~,
\end{equation}
where $0 \le \beta \le 1$.
The initial phase always winds by $2\pi$ around the whole ring.
When $\beta=0$, the initial phase winds linearly around the ring without any phase jump.
When $\beta=1$, the initial phase first increases linearly, then jumps by $\pi$ at $\tx=0$, and then continues to increase linearly such that the overall phase winding is still $2\pi$.
A maximum phase jump of $\pi$ is considered here because any phase jump $\Delta \theta > \pi$ is equivalent to a phase jump in the other direction with magnitude $2\pi - \Delta \theta$, and will correspond to an initial condition with no initial phase winding.
The two parameters $(d,\beta)$ now continuously connect two topologically different solutions: the constant-amplitude solution has $d=1$ and $\beta=0$, while the instanton solution has $d=-1$ and $\beta=1$.

In Sec.~\ref{subsec:region_of_instanton_formation}, we present the numerical results to show the region in this parameter space where the constant-amplitude or the instanton solution will arise.

\begin{figure}
    \centering
    \includegraphics[width=\columnwidth]{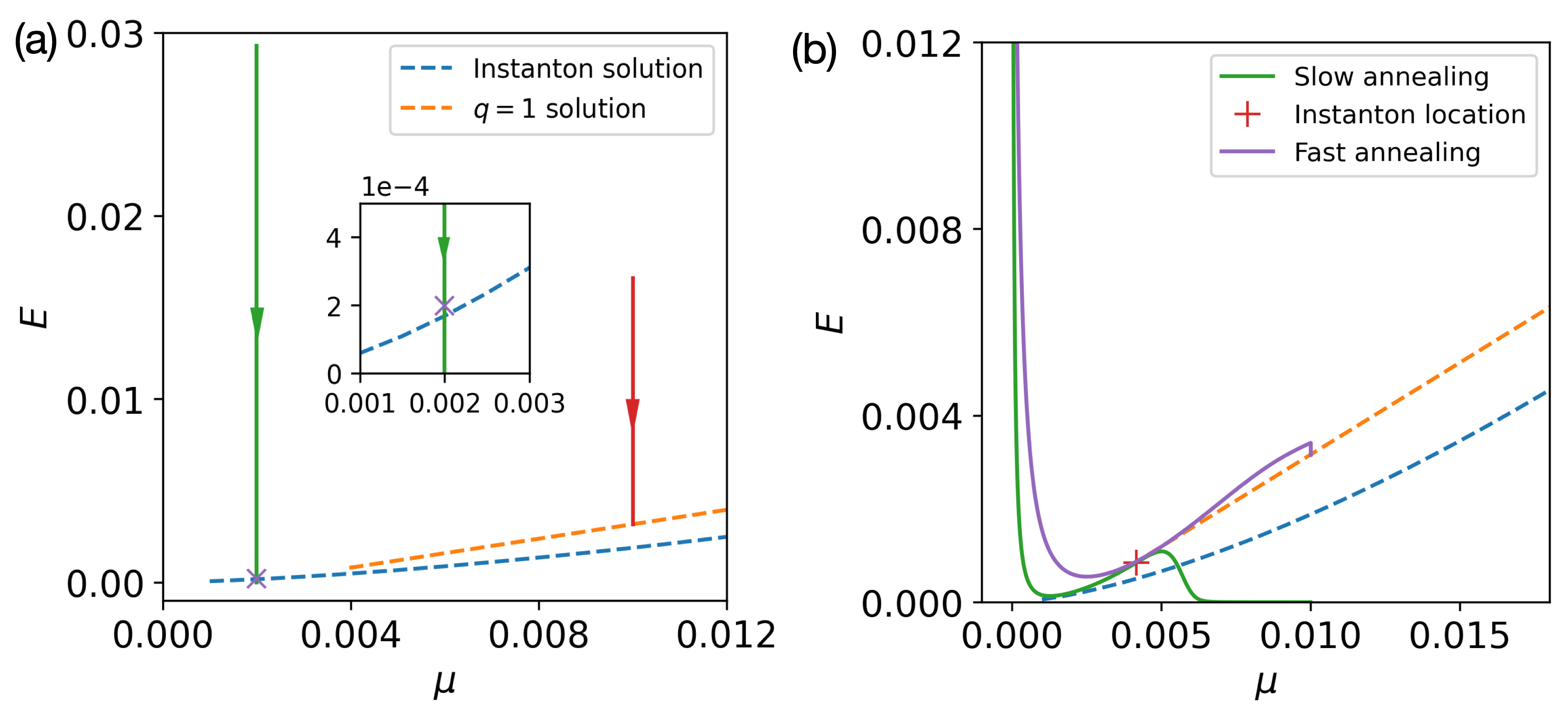}
    \caption{
   Trajectories during the dynamical evolution of Eq.~(\ref{eqn:continuous_pde}) on  the $E$–$\mu$ plane.  
\textbf{(a)} Fixed‑$\mu$ simulations from Fig.~\ref{fig:low_mu_evolution}:  
$\mu=0.002$ (green) falls below the $q=1$ branch (black dashed), nucleates an instanton (\textsf{\texttimes}) and relaxes to the ground state $E=0$;  
$\mu=0.01$ (red) lands on the $q=1$ excited branch (orange dashed).  
Inset: zoom confirms the instanton energy agrees with the analytic value from Sec.~\ref{subsec:analytical_ansatz}.  
\textbf{(b)} Linear ramps $\mu(t)\!:0\!\to\!0.01$ from the same initial state:  
slow anneal ($T=50$, solid) triggers an instanton (\textsf{\texttimes});  
fast anneal ($T=25$, dashed) does not.%
 }
    \label{fig:low_mu_energy_vs_mu}
\end{figure}

\begin{figure*}
    \centering
    \includegraphics[width=\linewidth]{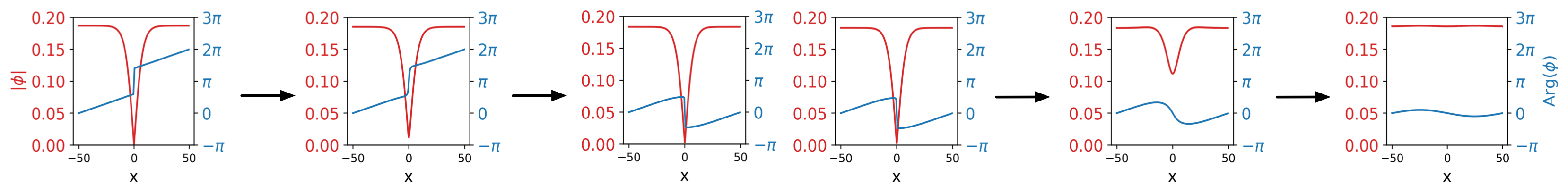}
    \caption{Evolution of $|\phi|$ (top) and $\arg\phi$ (bottom) for Eq.~(\ref{eqn:continuous_pde}) with fixed $\mu=0.04$.  
The initial amplitude matches the analytic instanton profile, while the phase is linear except for a $\tfrac45\pi$ jump at $x=100$, coinciding with the zero‑amplitude point.  
An instanton nucleates at this site, converting the initial twist to zero winding; panels 3–4 show the state immediately before and after the event.%
    }
    \label{fig:high_mu_evolution}
\end{figure*}

\section{Numerical Results}
\label{sec:numerical_results}
\subsection{Examples of Instanton Formation}
\label{subsec:examples_of_instanton_formation}
We initialize the ring with random amplitudes and a linear $0\!\to\!2\pi$ phase ramp.  
Figure~\ref{fig:low_mu_evolution} contrasts two numerical evolutions  of Eq.~(\ref{eqn:continuous_pde}) that differ only in the gain parameter~$\mu$: 
{\it Above threshold} ($\mu > (2\pi/L)^2$, panel (b)): the constant‑amplitude $q=1$ branch exists, so the energy decays until it reaches this excited branch and remains there.
{\it Below threshold} ($\mu < (2\pi/L)^2$, panel (a)): the $q=1$ branch is absent, so the energy keeps falling.  An instanton nucleates, removes the $2\pi$ twist, and the system relaxes to the defect‑free ground state.
Thus, a modest reduction in $\mu$ switches the long‑time attractor from the excited $q=1$ state to the true ground state via an instanton‑mediated phase slip.%

Figure~\ref{fig:low_mu_energy_vs_mu}(a) plots the  dynamical evolution according to Eq.~(\ref{eqn:continuous_pde}) on $(E,\mu)$ plane.
For large gain (red curve), the trajectory terminates on the $q=1$ branch, so $E$ cannot drop further.  
For small gain (green curve), the system slides beneath that branch; at the point marked \textsf{\texttimes} an instanton appears, the amplitude locally vanishes, a $2\pi$ phase slip occurs without an energy jump, and the trajectory reaches  $E=0$.  
No fine-tuned initial state is required because the instanton forms spontaneously from generic noisy data.%

The space–time maps in Fig.~\ref{fig:low_mu_evolution}(c,d) starkens the contrast.  
In the dynamical evolution for small $\mu$  [Fig.~\ref{fig:low_mu_evolution}(c)], the amplitude profile develops a local dip to $|\phi|=0$ and an accompanying $2\pi$ phase jump (an instanton) after which the twist rapidly vanishes.  
For the larger gain [Fig.~\ref{fig:low_mu_evolution}(d)], the amplitude never vanishes, no slip occurs, and the initial winding persists.
Crucially, the random initial state has spatial amplitude fluctuations; rather than averaging out, one of these depressions can deepen to zero and trigger the slip.

Many implementations vary the gain parameter in time (annealing) rather than keeping it fixed
\cite{yamamotoCoherentIsingMachines2017,inagakiCoherentIsingMachine2016,
kalininGlobalOptimizationSpin2018,kalininNetworksNonequilibriumCondensates2018}.
A slow ramp is known to boost the success rate of Ising/XY machines
\cite{syedPhysicsEnhanced2023,cumminsIsingHamiltonian2025}, and Fig.~\ref{fig:low_mu_energy_vs_mu}(b) confirms the same mechanism here.
Two dynamical evolutions of Eq.~(\ref{eqn:continuous_pde}) start from identical states and share the same initial and final $\mu$, but the slow schedule (total time 50) dwells longer in the low‑$\mu$ regime, allowing an instanton to nucleate; the fast schedule (total time 25) does not, and the system remains on the excited branch.
Hence, gentle annealing, by keeping $\mu$ small for longer, facilitates instanton formation and greatly increases the likelihood of reaching the ground state.

\begin{figure}
    \centering
    \includegraphics[width=\columnwidth]{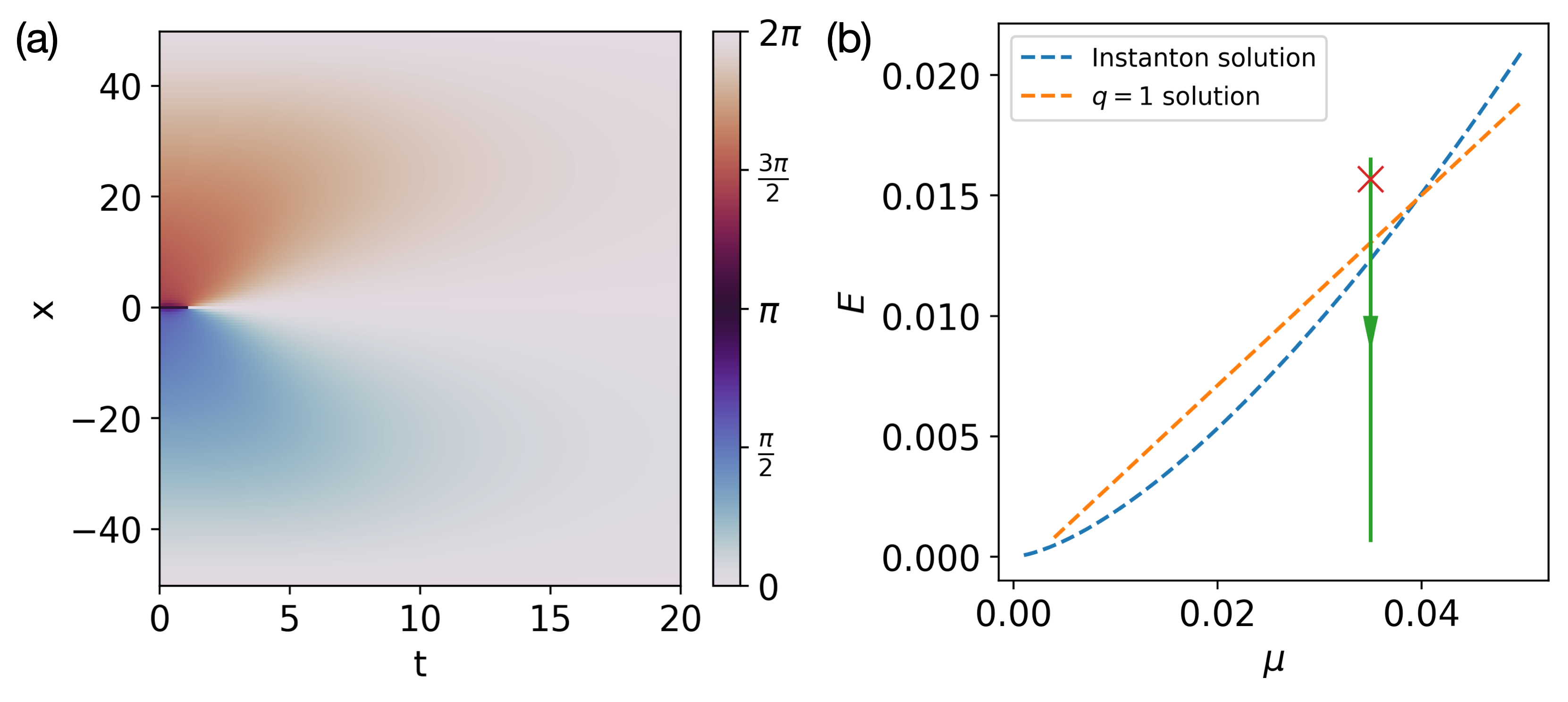}
    \caption{
    (a)~Space–time phase map for the run of Fig.~\ref{fig:high_mu_evolution}; an instanton forms shortly after $t=0$, removing the initial $2\pi$ twist.%
(b)~Energy trajectory for the same run (arrow indicates forward time).  The instanton event (\textsf{\texttimes}) bridges to the $E=0$ defect‑free ground state.%
}
    \label{fig:high_mu_phase_and_trajectories}
\end{figure}

All previous numerical simulations started from a near–uniform amplitude with a linear $2\pi$ phase ramp.  
Figure~\ref{fig:high_mu_evolution} explores a contrasting initial state: the amplitude is the instanton profile of Sec.~\ref{subsec:analytical_ansatz} (zero at one site), but the phase is linear everywhere except for an abrupt $\frac45\pi$ jump at that same site.  
This pre‑imposed amplitude dip seeds an instanton so effectively that it appears even at the much larger gain $\mu=0.35$ (third and fourth panels).  
The event links the $2\pi$‑wound state to the zero‑winding state without an energy barrier, because the phase slip occurs exactly where $|\phi|=0$.  
Notably, the dip first partially refills before collapsing to zero and triggering the $2\pi$ jump, illustrating that the instanton need not form immediately but is all but inevitable once a zero‑amplitude site is present.%

The evolution of the  phase distribution is shown in Fig.~\ref{fig:high_mu_phase_and_trajectories}(a), where we can clearly observe the formation of an instanton early in the dynamics.
From this point onward, phase variations across the ring start to smooth out, and the system eventually adopts the $q=0$ defect-free state.
From Fig.~\ref{fig:high_mu_phase_and_trajectories}(b), one can observe that the trajectory crosses the $q=1$ excited state without becoming trapped.
This is because the instanton is formed before its energy crosses the energy $q=1$ excited state, as shown by the location of the cross in Fig.~\ref{fig:high_mu_phase_and_trajectories}(b).
Anywhere on the trajectory below the cross, the system has no phase winding, so the system is separated from the excited state by a discrete non-zero topological charge.
This means that even as it passes through the $E(\mu)$ line given by the $q=1$ excited state, it is impossible for the system to become trapped into this excited state any more.

\subsection{Region of Instanton Formation}
\label{subsec:region_of_instanton_formation}
\begin{figure}
    \centering
    \includegraphics[width=\columnwidth]{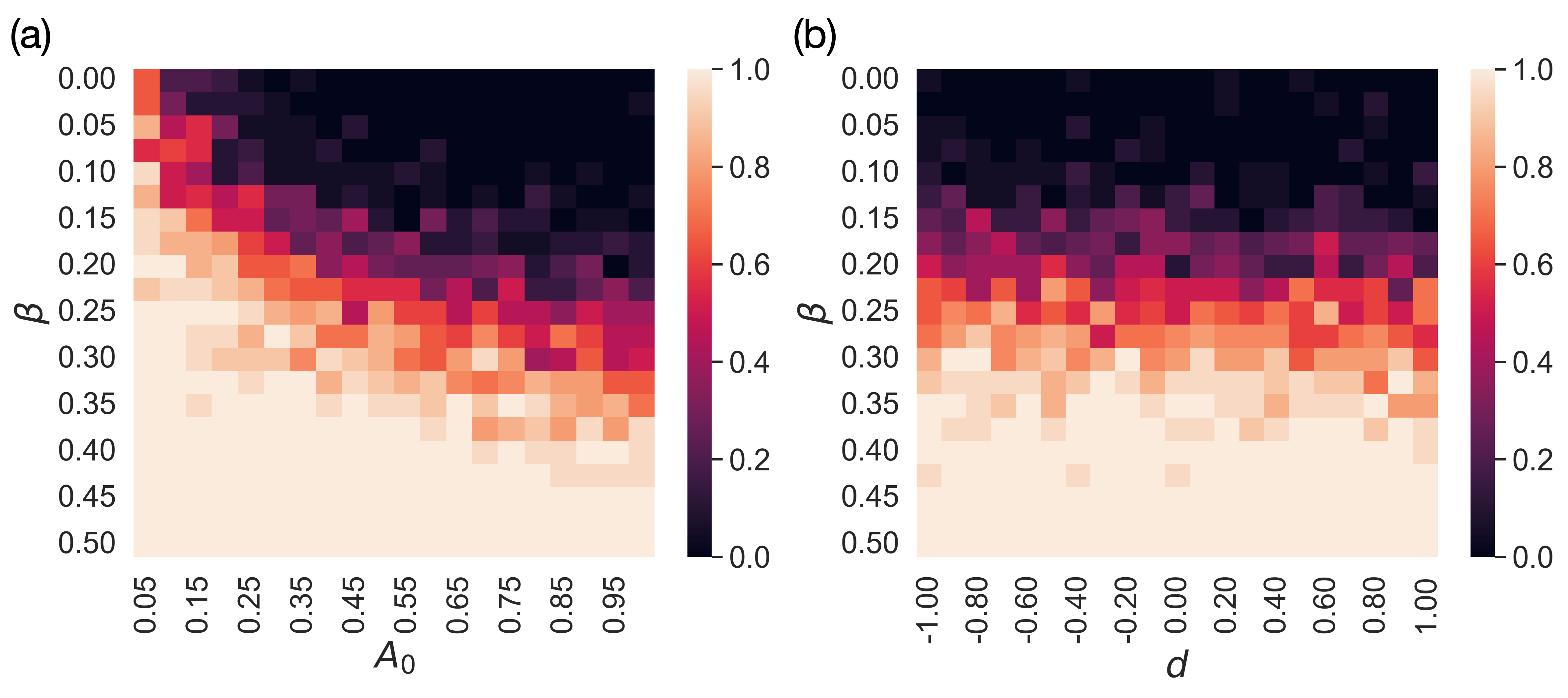}
    \caption{Probability of instanton formation for the scaled Eq.~(\ref{eqn:continuous_pde_scaled}) as a function of the parameters that define the initial state.  
(a)~Heat map in the $(\beta,A_0)$ plane with the dip‑depth fixed at $d=0$ (the amplitude falls to $A_0/2$ at $x=0$).  
(b)~Heat map in the $(d,\beta)$ plane with the overall amplitude factor fixed at $A_0=0.5$.  Here $d$ spans its entire admissible range, while $\beta$ is limited to $0\le\beta\le0.5$ because larger jumps always triggered an instanton.  
For each grid point 20 realisations were generated by adding uniform phase noise in $[-\pi/10,\pi/10]$ to the prescribed initial profile; colour encodes the fraction of runs that produced an instanton.  All simulations used a ring of scaled length $\tilde L=10$, corresponding to the unscaled $L$ employed in Sec.~\ref{subsec:examples_of_instanton_formation}.%
}
    \label{fig:init_cond_trend}
\end{figure}

Given that the two initial conditions considered above behaved very differently, we now seek to understand if the difference stems from the difference in the initial amplitude profile or the initial phase profile.
Using the parametrisation given in Eq.~\eqref{eqn:parameterised_amplitude} and Eq.~\eqref{eqn:parametrised_phase}, we can specify an initial condition by a set of parameters $(A_0, d, \beta)$, while making sure that the initial phase always winds by $2\pi$ around the whole ring.
$A_0$ and $d$ control the initial amplitude profile, while $\beta$ controls the phase profile.
When $\beta=0$, the initial phase winds linearly around the ring without any phase jump.
When $\beta=1$, the initial phase first increases linearly, then jumps by $\pi$, and then continues to increase linearly such that the overall phase winding is still $2\pi$.

Many initial conditions with different initial $(A_0, d, \beta)$ values were generated and simulated to observe if an instanton formed during their evolutions, which would untwist the initially present $2\pi$ phase winding and bring the system to the defect-free ground state.
For each set of $(A_0, d, \beta)$ values, $20$ random initial conditions were generated by adding a random noise uniformly distributed in the range $[-\pi/10, \pi/10]$ to the phase at each discretised location.
Each initial condition was evolved according to the scaled Eq.~\eqref{eqn:continuous_pde_scaled}, and the probability of instanton formation for each set of parameters is shown in Fig.~\ref{fig:init_cond_trend}.
In Fig.~\ref{fig:init_cond_trend}(a), $\beta$ and $A_0$ were varied while a constant value of $d=0$ was used.
It can be observed that the change in $\beta$, which quantifies the initial phase jump at the amplitude dip, significantly affects the probability of instanton formation.
Higher initial phase jump led to a higher probability of instanton formation, for any given $A_0$ value.
At the same time, it can also be observed that increasing initial amplitude level $A_0$ led to a narrower region of $\beta$ values in which instanton could form.
We note that the transition from the no-instanton to the instanton region does not have a sharp boundary.
The probability of formation of an instanton increased gradually with $\beta$ and decreased gradually with $A_0$.

In Fig.~\ref{fig:init_cond_trend}(b), $\beta$ and $d$ were varied while $A_0$ had a constant value of $0.5$.
The influence of $\beta$ on probability of instanton formation can still be observed in this plot, but in contrast, changing the initial amplitude dip depth, controlled by $d$, did not lead to a change in the probability of instanton formation.
Even at $d=-1$, where the initial amplitude dropped to $0$ at $x=0$, making the initial amplitude profile very close to that of an instanton, the probability of instanton formation remained almost the same as the case of $d=1$, where the initial amplitude was constant with no dip at all.

Hence, the combined results of Fig.~\ref{fig:init_cond_trend}(a) and (b) imply that an initial phase profile with larger phase jumps plays an important role in pushing the system towards the formation of an instanton during its evolution, while for the initial amplitude profile, the overall initial amplitude level, quantified by $A_0$ is a more important factor than the magnitude of amplitude variations, quantified by $d$.
However, we note that the presence of amplitude fluctuation in the dynamics remains a key ingredient in the formation of instanton and the unwinding of initial phase windings.

\section{Phase Unwinding via Rarefaction Fronts from Vortex--Antivortex Collisions}
\label{sec:twod}

We now extend our analysis to a \emph{two-dimensional toroidal} (doubly-periodic) domain.  As in the 1D case of instantons, amplitude freedom again proves crucial for removing a global $2\pi$ phase twist.  Here, the topological unwinding proceeds via vortex--antivortex collisions that create extended low-density ``rarefaction fronts.''  In what follows, we first clarify the topological constraint on a 2D torus and why a single vortex cannot remove the global phase winding (\ref{subsec:2d_topology}).  We then introduce \emph{rarefaction pulses} in the Ginzburg--Landau context (\ref{subsec:rarefaction_pulses}) and show how these appear transiently during vortex--antivortex annihilation (\ref{subsec:2d_vortex_collision}).  Finally, we contrast the small-$\mu$ regime, where large vortex cores merge into a domain-spanning front, with the large-$\mu$ regime, where winding persists (\ref{subsec:2d_fail_unwind}).

\subsection{Topological Winding on a 2D Torus}
\label{subsec:2d_topology}

On a doubly-periodic domain $(x,y)\in [0,L_x]\times[0,L_y]$, a net $2\pi$ phase twist can be imposed along the $x$-direction by setting
\[
\arg\!\bigl(\psi(x+L_x,y,0)\bigr) 
\;=\; 
\arg\!\bigl(\psi(x,y,0)\bigr) \;+\; 2\pi q.
\]
Such a global winding $W_x=q$ cannot be continuously unwound if $|\psi|>0$ everywhere, as it is topologically locked.  A single vortex in 2D  has a $2\pi$ phase circulation around its core but does not alter the boundary condition around the entire horizontal cycle.  Hence, removing a net winding requires two oppositely charged vortices (a vortex-antivortex pair) whose annihilation can cut a zero-density path across the domain, effectively letting the phase slip by $2\pi$.

For the dissipative Ginzburg--Landau equation, Eq.~\eqref{eqn:continuous_pde}
the parameter $\mu$ sets the background amplitude $\sqrt{\mu}$ and the \emph{healing length} $\xi\!\sim\!1/\sqrt{\mu}$, which characterizes vortex-core size. The vortex  is a stationary solution of Eq.~\eqref{eqn:continuous_pde} given by $\phi(r)=R(r) e^{\pm i \theta}$ with amplitude satisfying 
\begin{equation}
R''+ \frac{R'}{r}   - \frac{R}{r^2} + (\mu - R^2)R=0,
\label{vortex}
\end{equation}
with boundary conditions $R(0)=0, R(\infty)=\sqrt{\mu}.$ The vortex core (the characteristic length on which the vortex amplitude heals itself) is $\xi\sim 1/\sqrt{\mu}.$
Pad\'e approximations are helpful in finding an approximate  expression for the vortex core with the correct asymptotics at the centre of the vortex core and away from it \cite{berloff2004pade}, and for a general $\mu$ can be found as
\begin{equation}
   \psi_{v} = \sqrt{\mu}r \sqrt{\frac{a_1 \mu + a_2 r^2}{\mu^2 + b_1 \mu + a^2 r^4}} \exp[\pm i \theta],
\end{equation}
where $a_1\approx 0.3437, b_1 \approx 0.3333,$ $a_2\approx 0.0286$.
  When $\mu$ is small, each vortex has a broad core, permitting two nearby vortices to merge their low-density regions into one connected front.  Conversely, a large $\mu$ yields small, point-like vortices that never coalesce into a domain-spanning rarefaction wave.  This difference underlies the success or failure of phase unwinding in 2D, as we will show.

\subsection{Rarefaction Pulses in the Ginzburg--Landau Context}
\label{subsec:rarefaction_pulses}

Although ``rarefaction pulses'' were derived initially as \emph{stable solitary waves} in \emph{conservative} nonlinear Schr\"odinger (NLS) equations \cite{jones1982motions,berloff2002evolution,berloff2004interactions, berloff2004motions, berloff2004pade, pinsker2014transitions}, they also arise in our gain-dissipative Ginzburg--Landau framework as transient or saddle-point configurations formed by vortex--antivortex collisions.  In a Hamiltonian BEC, a rarefaction pulse (or Jones--Roberts soliton) is a vortex-free density dip travelling at high velocity, belonging to a continuous solitary-wave branch whose low-velocity limit is a vortex dipole \cite{jones1982motions,neely2010observation,meyer2017observation,baker2025observation}.  When the flow speed is above a critical value, the vortices merge into a single dark wave  (the rarefaction pulse); below that speed, the same solitary wave separates into two discrete vortex cores.

In a purely conservative environment at zero temperature, such pulses can propagate indefinitely as stable solitary waves.  By contrast, under our dissipative Ginzburg-Landau evolution, these same local density profiles appear only transiently as the system relaxes.  They serve as short-lived saddle points, so once a $2\pi$ phase slip is enacted, the system continues dissipating to the uniform ground state.  

{\it Mechanism for Defect Annihilation.}
Despite not forming a stable solitary wave, these transient pulses accomplish a crucial topological task: creating an extended region of $|\psi|\approx 0$ that cuts across the torus, enabling a global $2\pi$ phase flip.  In simpler phase-only models, the amplitude is fixed, and no analogous low-density channel can exist, so the system remains locked in the winding sector.  Hence, the amplitude freedom in Ginzburg-Landau is the key enabler, and rarefaction pulses provide a direct 2D analogue of the 1D instanton-mediated phase slip.

\subsection{Collision of Large-Core Vortex--Antivortex Pairs at Small \(\mu\)}
\label{subsec:2d_vortex_collision}

Concretely, consider a random initial condition with a net winding $W_x=1$.  Under gradient-flow dynamics, forming a vortex-antivortex pair lowers the system’s free energy if $\mu$ is small enough that each vortex core is large and easily overlaps with the other.  As these two opposite vortices collide or coalesce, they generate a domain-spanning channel of suppressed amplitude -- a “sheet” reminiscent of a rarefaction pulse.

\begin{itemize}
\item \textbf{Broad core overlap.}  Large healing length $\xi \sim 1/\sqrt{\mu}$ means each vortex has an extended region of depressed $|\psi|$.  When oppositely charged vortices come together, their cores merge, forming a continuous low-density line.
\item \textbf{Global phase slip.} At or near the moment of annihilation, the part of the channel hits $|\psi|\approx 0$, allowing the phase to jump by $2\pi$ in that part, seeding the removal of  the winding as a secondary pair of vortex-antivortex propagates across the channel, completing the phase unwinding.  We, therefore, see a global topological change.
\item \textbf{Transient rarefaction front.}  Although short-lived, the geometry is indeed that of a vortex-free dark soliton (i.e.\ rarefaction pulse).  Once the unwinding occurs, the final amplitude recovers toward $\sqrt{\mu}$ everywhere, relaxing to the ground state $W_x=0$.
\end{itemize}

\begin{figure}
    \centering
    \includegraphics[width=\columnwidth]{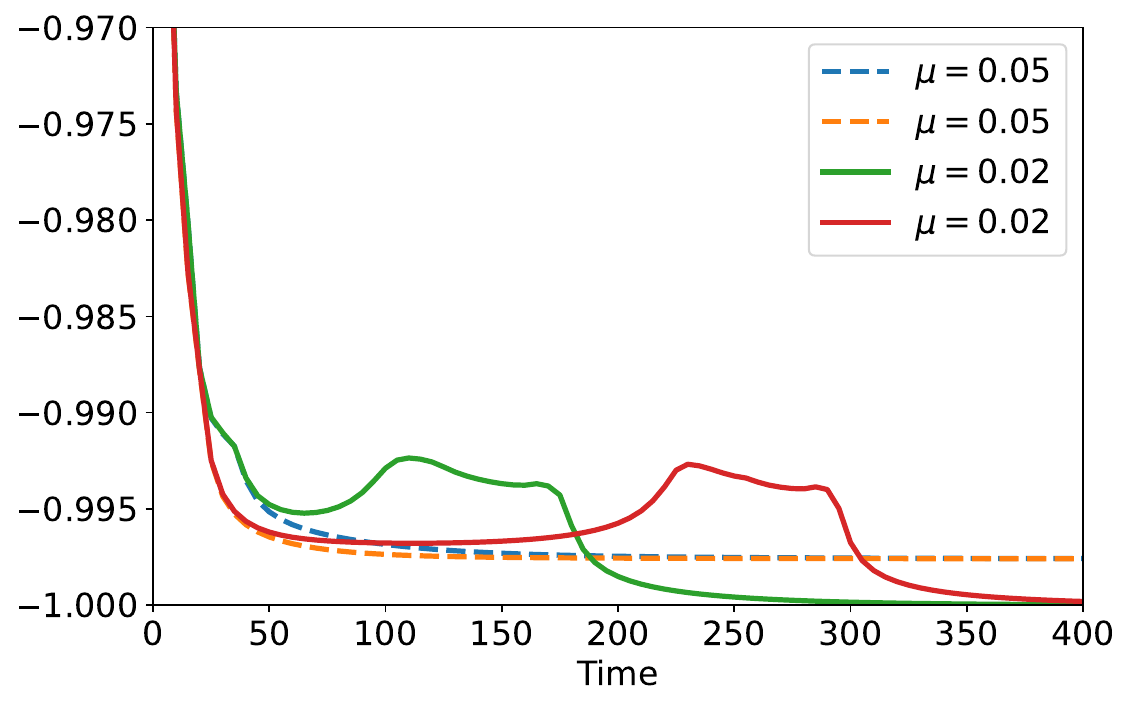}
    \caption{XY energy during $64\times64$ Stuart–Landau simulations starting from two different random conditions and two different gains.  
The Hamiltonian $H_{\mathrm{XY}}=\tfrac1{N^2}\sum_{i,j}\cos(\theta_i-\theta_j)$ is tracked while Eq.~(\ref{eqn:sl_complex}) is integrated with $dt=0.05$, $J_{nm}=1$, and either $\mu=0.02$ or $\mu=0.05$ (legend).  Both runs begin from the same random state with a horizontal winding $W_x=1$.  For $\mu=0.05$, the energy plateaus on the excited $W_x=1$ branch; for $\mu=0.02$, it continues downward to the $W_x=0$ ground state.%
}
    \label{fig:twod_energy.vs.time}
\end{figure}

Figure ~\ref{fig:twod_energy.vs.time} shows an illustrative example starting with the same initial condition:  at small $\mu=0.02$, the system successfully unwinds $2\pi$ winding, for larger $\mu=0.05$  the system evolves into the excited state with $W_x=1$. The energy of the system is always decreasing with time, but to gain an understanding of the evolution, we instead follow the energy of the XY Hamiltonian. For larger $\mu$, the XY Hamiltonian simply relaxes to the excited state, whereas for smaller $\mu$, to make the phase unwinding, the system has to go over the energy barrier of the XY Hamiltonian to reach the ground state and the amplitude dynamics allows this.  In Fig.~\ref{fig:twod_contourplots}, we see that the rarefaction front for smaller $\mu$ is sufficiently deep (near-zero amplitude) to effect the phase slip.  By $t=265$, the system has no net winding and converges to the uniform solution.

\subsection{Persistence of Winding at Large \(\mu\)}
\label{subsec:2d_fail_unwind}

\begin{figure}
    \centering
    \includegraphics[width=\columnwidth]{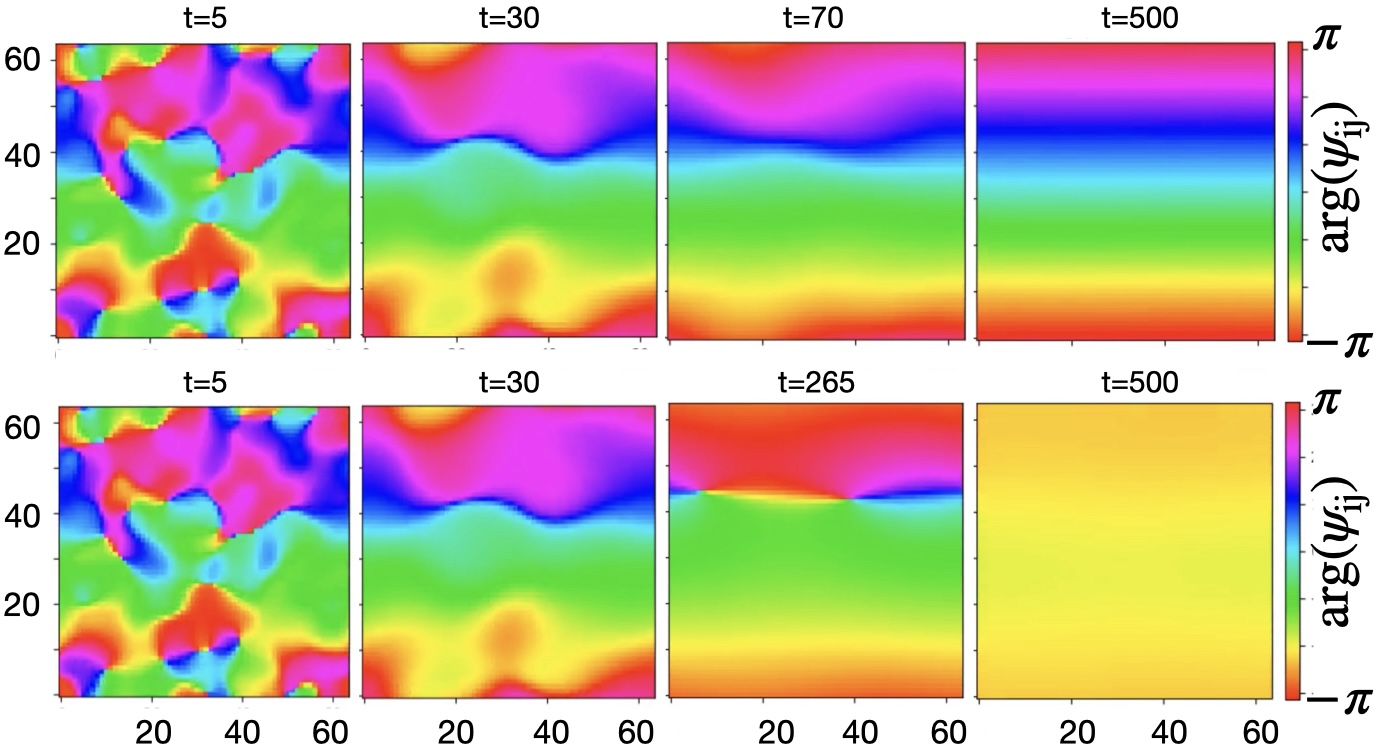}
    \caption{Phase snapshots for a $64\times 64$ Stuart–Landau lattice [Eq.~(\ref{eqn:sl_complex})] integrated with $dt=0.05$ and lattice spacing $l=1$.  
Top row: $\mu = 0.05$; bottom row: $\mu = 0.02$.  Both runs start from the same random state with a horizontal winding $W_x=1$.  
At $t=5$, a vortex–antivortex pair appears; by $t=30$, their annihilation produces a rarefaction pulse.  For $\mu=0.05$ the pulse is too shallow to trigger a phase slip, and the system stabilises in the excited $W_x=1$ state.  For $\mu=0.02$, the pulse deepens to zero amplitude, enabling a global phase slip: two defects propagate vertically ($t=265$) and the lattice relaxes to the $W_x=0$ ground state.%
}
    \label{fig:twod_contourplots}
\end{figure}

Conversely, if $\mu$ is large, each vortex core is small and expensive to create or expand.  Even if a vortex--antivortex pair appears, their annihilation remains localised—no spanning “rarefaction pulse” emerges.  Hence, no global $2\pi$ slip is possible, and the system remains in a \emph{metastable current state}.  Figure~\ref{fig:twod_contourplots}  exemplifies this scenario: for $\mu=0.05$, the vortices form but never produce a domain-wide amplitude depression, so the final state retains $W_x=1$.  

In sum, amplitude freedom again underpins phase unwinding: for small $\mu$, large-core vortices overlap into a transient rarefaction wave that cuts across the domain, permitting a global topological slip.  At large $\mu$, the vortices remain localised, no extended low-density channel appears, and the system stays trapped in the winding sector.  This parallels the 1D situation (instanton vs.\ no slip) but is realized here through vortex--antivortex collisions and rarefaction pulses -- the true 2D analog of dark soliton flips.

\section{Discussion and Conclusions}
\label{sec:discussion_conclusion}
The present work identifies an explicit mechanism via the formation of an instanton (in 1D) or rarefaction pulses (in 2D) that enables the unwinding of global phase twist in an optical network. While the qualitative idea that amplitude freedom allows defect healing has long been recognized, previous studies typically framed the phenomenon in either broad heuristic terms or in context-specific numerical observations. By contrast, we identify and analyze in detail an analytical solution within a gain-dissipative Stuart-Landau framework, concretely demonstrating that local amplitude collapse to zero enables the system’s phase to “slip” by $2\pi$ without incurring a jump in energy. This mathematical analysis shows that the presence of  instantons and rarefaction pulses effectively bridges different integer winding sectors in a continuous trajectory. Our numerical simulations further establish that this phenomenon is robust to noise and a range of initial conditions, explaining why amplitude-driven oscillators can avoid getting trapped in excited states with nonzero winding.

In this work, we uncovered a unifying amplitude-driven mechanism for removing global phase windings in the one-dimensional ring and two-dimensional toroidal oscillator networks governed by gain-dissipative dynamics.  Our analysis reveals two primary levers for guaranteeing that any initially present topological defects unwind, thus allowing the system to reach its fully synchronized ground state:

\textbf{1. Operating at low effective injection (just above threshold).}  
When the linear drive $\mu$ (or injection/pump parameter) is kept close to but above the onset of lasing, the system remains “soft” in amplitude.  Consequently, local collapses to near-zero amplitude occur more readily, enabling phase-slip events in 1D (instanton formation) or vortex--antivortex collisions in 2D that produce extended rarefaction pulses.  If $\mu$ is taken too large, vortices become small-core objects that cannot extend across the domain or create a system-spanning front, thereby locking in the initial winding.  Thus, lower $\mu$ fosters amplitude suppression and simplifies global phase unwinding.

 \textbf{2. Choosing or engineering initial conditions that favor amplitude dips.}  
Even with moderately large $\mu$, it is still possible to form instantons or rarefaction pulses if the initial amplitude and phase profiles promote the development of local zero-amplitude regions.  For instance, introducing a well-placed phase jump or an amplitude trough at specific sites can effectively seed the topological unwinding process.  Conversely, uniform or high-amplitude initial configurations may trap the system in excited states with nonzero winding.  Hence, adopting tailored initial states accelerates defect removal.

Our numerical sweeps confirmed that both a low injection rate and/or a favourable initial condition could facilitate the amplitude collapse at a minimal energy cost, driving the system across topological barriers.  Related work on “bifurcation-based” or “annealing-based” optical solvers \cite{inagakiCoherentIsingMachine2016, syedPhysicsEnhanced2023} similarly finds that slowly raising the effective gain leaves time for amplitude fluctuations to remove local minima—consistent with our conclusion that spending longer in a low-$\mu$ regime improves the probability of eliminating phase windings.

{\it Implications for analog solvers.}
From an optimization standpoint, persistent topological defects can represent false minima, preventing networks from discovering the true ground state.  Our findings demonstrate that amplitude freedom is indispensable for escaping these traps: phase-only models (e.g.\ Kuramoto) cannot shed global windings, while Stuart–Landau or laser-rate equations readily do so via zero-amplitude cores.  Practically, this implies that photonic or oscillator-based hardware aiming to solve XY-type spin problems should: A. Maintain pump levels only slightly above threshold, ensuring a greater range of amplitude fluctuations;
B. Introduce mild inhomogeneities or “phase jumps” in the initial condition so the system does not begin in a uniform high-amplitude state resistant to phase slips;
C. Consider slow ramping protocols (“annealing”) that preserve low amplitude for enough time to let instantons or rarefaction fronts nucleate and remove windings.

{\it Outlook.}
Although we focused primarily on ring lattices (1D) and toroidal grids (2D) and ferromagnetic couplings, the same principles hold for more general coupling graphs—where amplitude collapses can remove loop windings or vortex-like defects.  Extending these insights to larger dimension or to spin-glass–type connectivity remains a compelling avenue, as does integrating active feedback to strategically trigger amplitude dips.  We envision that instanton-aided photonic optimization devices could combine amplitude control, targeted seeding, and slow injection schedules to systematically avoid local minima.  
In short, keeping the injection rate barely above the threshold, along with ensuring a non-uniform initial condition, emerges as a potent recipe for global phase ordering in gain-dissipative oscillator networks.  This amplitude-enabled unwinding (instantons in 1D or rarefaction pulses in 2D) provides a promising framework for next-generation photonic solvers, where robust and scalable computation depends critically on reliably removing topological defects.

\section*{Acknowledgements}
R.Z.W. and N.G.B.\ acknowledge the support from HORIZON EIC-2022-PATHFINDERCHALLENGES-01 HEISINGBERG Project 101114978 and the support from the Julian Schwinger Foundation Grant No.~JSF-19-02-0005.
N.G.B.\ acknowledge support from Weizmann-UK Make Connection Grant 142568.

\bibliography{Instanton_References}
\end{document}